# An Approach for Design Parameter Optimization of the Triangle Cloud Control System

UnSun Pak, YongNam Sin, GyongIl Ryang

Faculty of Electronics & Automation, **Kim Il Sung** University, Pyongyang,

Democratic People's Republic of Korea

**Abstract:** In this paper, we have proposed the optimization approach of design parameter of generalized cloud control system by hybrid chaos optimization approach.

The approach determined the off-line parameters of the cloud control system by the chaos approach and on-line by gradient approach.

## 1. Introduction

In this paper, an approach for optimization of the Cloud Control System which is a part of Intelligent Control had been proposed.

In previous research [1-10], the design approach of the Cloud Controller had only been proposed, and the methods for optimization of the Control System had not been proposed and an approach that optimizes the parameters of the Fuzzy Controller with the Chaos Optimal Method had been proposed but this method has a disadvantage of which is difficult to apply the plant whose variance is awful because this method optimizes the parameters of the controller on off-line.

In paper, an approach that optimizes the parameters of the Triangle Cloud Controller with the hybrid chaos approach which is combined with the Chaos Optimization Method and the Gradient Method has been proposed.

## 2. The off-line optimization of the design parameter of the triangle cloud control system by the parallel variable scaling rate chaos optimization approach

First, Using the Hybrid Chaos Optimal Method with the parallel variable chaos optimal method and the conjugation gradient method, the system configuration figure for optimizing the design parameters of the triangle cloud controller is shown fig.1.

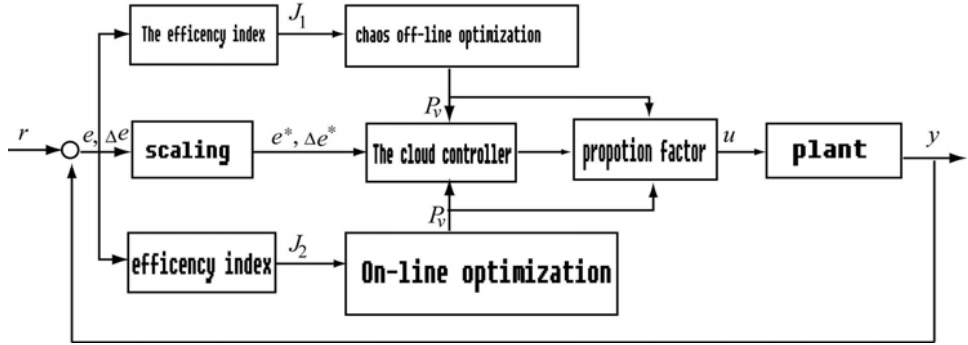

Fig.1 System configuration figure for design optimization of the cloud controller

As you're seen on figure, the triangle cloudy controller is a 2-input-1-output system construction and the front membership cloudy function is the triangle member cloudy function, the back member function is single point member cloud function.

Creating the expected value, variance, super variance, division number of the front member triangle cloudy function, and the expected value, division number and cloudy rule is decided on off-line that the efficiency index J1 gets the minimum with the parallel variable reducing rate chaos optimal method.

And then we decide the parameters of the cloudy controller on on-line that the efficiency index J2 gets minimum with the conjugation gradient method.

Using parallel variable reducing rate chaos optimization method, the efficiency index J1 is expressed like this.

$$J_1 = \min[\sum_{k=1}^{T} k^2 |e(k)| \Delta t]$$

This optimizes the control parameters with parallel variable scaling rate chaos optimal approach

$$J_1 = \min[\sum_{k=1}^{T} k^2 |e(k)| \Delta t] \qquad (1)$$

This expression means that the error area gets minimal all over the control interval.

In the expression, $e(k) = r(k) - y(k)$ means error, $\Delta t$ means sampling interval, $r(k)$ means the quantity of the target institution, $y(k)$ is the output of the plant.

And then let express the parameters; $Ex_{x_1}, Ex_{x_2}$, - the expected value, $En_{x_1}, En_{x_2}$ -the variance, $He_{x_1}, He_{x_2}$ -the super variance of the front member cloud function and $Ex_u$ -the expected value of the after member cloud function, the cloud division number of the first input variable of the cloud controller, the cloud division number of the second input variable, the division number of the after member cloud function, creating the cloud rules and the proportion coefficients as $P_v, v = 1, 2, \cdots$ that is

$P_1 = Ex_{x_1}, P_2 = Ex_{x_2}, P_3 = En_{x_1}, P_4 = En_{x_2}, P_5 = He_{x_1}, P_6 = He_{x_2}, P_7 = Ex_u, P_8 = m_1, P_9 = m_2,$
$P_{10} = o, P_{11} = RL, P_{12} = K_u$

Here $P_1, P_2, P_3, P_4, P_5, P_6, P_7$ are the parameters of the front and the after member cloud function so in the case of the division number is $P_5, P_6, P_7$, they become as

$P_1^i, P_3^i, P_5^i, P_2^j, P_4^j, P_6^j, P_7^h, i = m_1, j = m_2, h = o$ and

$P_8 = RL = [RL^1, \cdots, RL^{\kappa}, \cdots], \kappa = m_1 \times m_2$.

And then $P_1, P_2, P_7$ have the value of from -1 to 1, $P_3, P_4, P_5, P_6$ have the value of from 0 to 1, $P_8, P_9, P_{10}$ have the value of 1 to 20, $P_{11}$ has the value of 1 to o which is the division number of the after member cloud function, $P_{12}$ has the value of $P_u = \max(|u_{\min}|, |u_{\max}|)$.

Now let's study about Logistic mapping- $\alpha_{s+1} = 4\alpha_s(1-\alpha_s), s = 1, 2, \cdots, N, \alpha_0 \in (0,1)$

to use the parallel variable scaling chaos optimal approach
In this expression N is the degree of the chaos optimization.

The chaos variable $\alpha_s$ takes the value of from 0 to 1, so we must modify $P_v$ as the following expression to determine the parameters of the cloud controller.

$$P_1, P_2, P_7 = -2\alpha + 1$$
$$P_3, P_4, P_5, P_6 = \alpha$$
$$P_8 = P_9 = P_{10} = round(20 \times \alpha)$$
$$P_{11} = round(o \times \alpha)$$
$$P_{12} = round(P_u \times \alpha) \quad (3)$$

The optimal process for which determine the parameters of the triangle cloud controller is following.

Step1. In the cloud controller, the number of parameter which will be determined is following.
$\gamma = m_1 \times 3 + m_2 \times 3 + o + 4 + m_1 \times m_2$

Take the different forty-one initial random values in the interval between 0 and 1.

Step2. The initial values that are gotten are substituted to expression (2) and then we can get the $\gamma$ chaos trajectory variables.

Step3. Using expression (3), calculate the control force by substituting modified $\gamma$ variable to the output cloud controller and then calculate the efficient index J1 based on the expression (1) by substituting to the system model.

Step4. If the termination condition is satisfied, you should complete searching and go to step5, but if it isn't, you should return to setp2.

Step5. Look for the minimal value of the system efficient index and

Put out the optimal solution at that time.

## 3. On-line optimization of the design parameter of the Triangle Cloud Controller with the Conjugation Gradient Approach

If you optimize the parameters of the controller on off-line with the previous chaos search approach, you couldn't react actually about the random parameter variance.

So we can determine the parameters approximately with the chaos search approach and should optimize the parameters of the controller on on-line with the gradient approach by using them as initial value.

For this, we should decide the efficient index as the following expression.

$$J_2 = \frac{1}{2}e(t)^2, \quad e(t) = r(t) - y(t) \tag{4}$$

The negative gradient direction of the parameter vector about efficient index J2
$P = [P_1, P_2, P_3, P_4, P_5, P_6, P_7, P_8, P_9, P_{10}, P_{11}, P_{12}]^T$ is

$$W' = -\nabla J_2(P) = \sum_{k=1} e_k \frac{\partial y_k}{\partial P_k} = \sum_{k=1}(r_k - y_k)\frac{\partial y_k}{\partial P_k} \tag{5}$$

And the step width $\eta$ of the optimal search direction is determined as the following expression.

$$\frac{dJ_2(x_{i,k} + \eta W'_k)}{d\eta} = 0 \tag{6}$$

The correction expression of the parameter vector which we are going to obtain is
$P_{9,k+1} = P_{9,k} + \eta_k \cdot W'_k$.

Here, $P_{i,k}$ is the ith parameter vector in the kth iteration, $W'_k$ is the search direction vector in the kth search.

The algorithm of the conjugation gradient is

$$W'_k = -\nabla J_2(P_{i,k}) + \delta_{k-1} \cdot W'_{k-1} \tag{7}$$

And the conjugation coefficient of the (k-1)th step is

$$\delta_{k-1} = \frac{[\nabla J_2(P_{9,k})]^T \cdot [\nabla J_2(P_{9,k})] - [\nabla J_2(P_{9,k})]^T \cdot [\nabla J_2(P_{9,k-1})]}{[\nabla J_2(P_{9,k-1})]^T \cdot [\nabla J_2(P_{9,k-1})]}.$$

The process for which determines the parameters of the cloud controller with the conjugation gradient is following.

Step1. The $\gamma$ optimal solutions which is gotten by searching with the parallel variable scaling chaos optimal approach is substituted to the expression (5) as the initial value.

Step2. Calculate the step width $\eta$ of the optimal search direction by using expression (6).

Step3. Calculate the efficient index J2 by substituting the new $\gamma$ variables as the parameters of the cloud controller, which are calculate by being based on expression (7).

Step4. Substitute $P_{9,k+1}$ as $P_{9,k}$ and return to step1.

## 4. Experiment

Now, let's see the plant is express as the following discrete system equation.

$$y(k) = 3.737\,y(k-1) - 4.212\,y(k-2) + 1.492\,y(k-3) + 0.17u(k-1) - \\ - 0.238u(k-2) + 2.94u(k-3) + \delta(k)$$

Here $\delta(k)$ is the white noise whose mean value is zero and variance is 1.

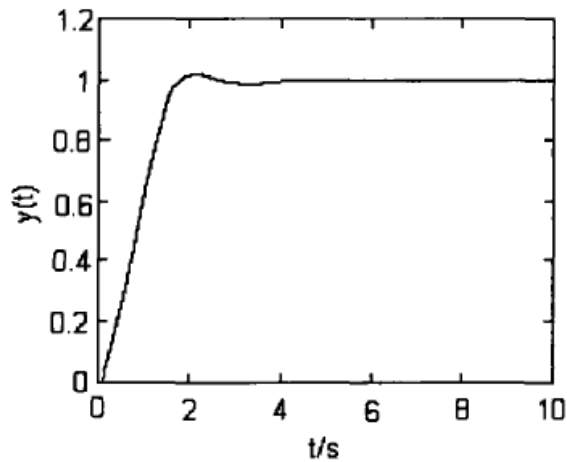

Fig.2 The response curve line of the system

First, Fig2 shows the result that are obtained by searching the global optimal solution accorded with the efficient index J1 on 837 steps with the chaos optimal approach and then obtaining the solution accorded with the efficient index J2 on 261 steps with the gradient approach and finally doing cloud control.

When fix the search completion as $J = 10^{-3}$ and optimize the cloud controller with the single chaos optimal method, the conjugation gradient descent method, genetic algorithm and the hybrid chaos optimal method, the frequency of obtaining the optimal solution is following.

Table. The degree of obtaining the optimal solution with the different optimal methods

| No. | The optimization approach | The degree of finding the optimal solution |
|---|---|---|
| 1 | The single chaos optimization method | 1539 |
| 2 | The conjugation gradient descend method | ∞ |
| 3 | Genetic algorithm | 1651 |
| 4 | The hybrid chaos optimization method | 1098 |

From the table, the times on which obtain the optimal solution with the conjugation gradient is ∞ so this method is easy to fall into the limit minimum value, you could notice that the global optimal solution can't be obtained.

And you can notice that the single chaos optimal method and the genetic algorithm are the global chance search method so we can find the global solutions with them and although the single chaos optimal method is more predominate than the genetic algorithm, they are not so good than the proposed hybrid chaos optimal method.

You can also notice that this regulator has a disturbance restraint character from the system response trajectory.

## 5. Conclusion

First, the approach which optimizes the design parameters of the triangle cloud control system on off-line with the parallel variable scaling chaos optimization method had been proposed.

Next, the approach which optimizes the design parameters of the triangle cloud control system with the gradient method had been proposed. Next, the efficiency of the proposed approaches had been verified through the computer simulation experiment.